\documentclass[journal]{IEEEtran}
\usepackage{cite}
\usepackage[numbers]{natbib}
\usepackage{amsmath,amssymb,amsfonts}
\usepackage{graphicx}
\usepackage{subcaption}
\usepackage{color}
\usepackage{multirow}
\usepackage{amsmath}
\usepackage{booktabs}
\ifCLASSINFOpdf
\else
\fi
\begin{document}
%
\title{Hybrid Data Augmentation and Deep Attention-based Dilated Convolutional-Recurrent Neural Networks for Speech Emotion Recognition}
%
%
%

\author{Nhat~Truong~Pham,
		Duc~Ngoc~Minh~Dang,
        and~Sy~Dzung~Nguyen
\thanks{This research was funded by the Vietnam National Foundation for Science and Technology Development (NAFOSTED) under grant number 107.01-2019.328. \textit{(Corresponding author: Sy Dzung Nguyen.)}}
\thanks{Nhat Truong Pham and Sy Dzung Nguyen are with the Division of Computational Mechatronics, Institute for Computational Science, Ton Duc Thang University, Ho Chi Minh City, Vietnam; Faculty of Electrical and Electronics Engineering, Ton Duc Thang University, Ho Chi Minh City, Vietnam (e-mails: phamnhattruong.st@tdtu.edu.vn; nguyensydung@tdtu.edu.vn).}
\thanks{Duc Ngoc Minh Dang is with School of Graduate Studies, Ton Duc Thang University, Ho Chi Minh City, Vietnam; Faculty of Electrical and Electronics Engineering, Ton Duc Thang University, Ho Chi Minh City, Vietnam (e-mail: dangngocminhduc@tdtu.edu.vn).}
}

\maketitle

\begin{abstract}
Speech emotion recognition (SER) has been one of the significant tasks in Human-Computer Interaction (HCI) applications. However, it is hard to choose the optimal features and deal with imbalance labeled data. In this article, we investigate hybrid data augmentation (HDA) methods to generate and balance data based on traditional and generative adversarial networks (GAN) methods. To evaluate the effectiveness of HDA methods, a deep learning framework namely (ADCRNN) is designed by integrating deep dilated convolutional-recurrent neural networks with an attention mechanism. Besides, we choose 3D log Mel-spectrogram (MelSpec) features as the inputs for the deep learning framework. Furthermore, we reconfigure a loss function by combining a softmax loss and a center loss to classify the emotions. For validating our proposed methods, we use the EmoDB dataset that consists of several emotions with imbalanced samples. Experimental results prove that the proposed methods achieve better accuracy than the state-of-the-art methods on the EmoDB with 87.12\% and 88.47\% for the traditional and GAN-based methods, respectively.
\end{abstract}

%
\IEEEpeerreviewmaketitle

\section{Introduction}
\IEEEPARstart{S}{peech} emotion recognition has a significant role in a lot of applications like e-learning, healthcare, HCI, customer relationship management (CRM), robotics, and video games~\cite{yoon2007study}--\cite{cen2016real}. In the past decades, SER has been one of the hot research topics in the speech processing areas. Most of previous works tried to use different type of features for the SER, such as pitch, energy, zero-crossing rate (ZCR), formants, root mean square error (RMSE), prosodic~\cite{cao2015speaker}--\cite{grimm2007primitives}, Mel-frequency cepstrum coefficients (MFCC)~\cite{el2007speech}--\cite{albornoz2011spoken}, linear predictive coding (LPC) and log frequency power coefficients (LFPC)~\cite{yeh2011segment}. Besides, the researcher tried to reach a variety of classification schemes to classify or discriminate the emotions like hidden Markov model (HMM), Gaussian mixture model (GMM), support vector machine (SVM), k-nearest neighbors (K-NN), and Bayesian logistic regression~\cite{cao2015speaker}--\cite{wu2011automatic}. Thanks to the development of deep learning, deep neural networks (DNN) have been used to automatically extract features for the SER instead of handcrafting~\cite{han2014speech}--\cite{huang2019feature}. Practically, deep convolutional neural networks (CNN) and long short-term memory (LSTM) have been successfully used to extract and exploit the time-frequency domain-based features from spectrograms~\cite{satt2017efficient}--\cite{issa2020speech}. The other studies applied an attention mechanism to the SER to obtain the most utterance features that represent the corresponding emotion~\cite{chen20183}--\cite{peng2020speech}. However, using the hierarchical and complex deep learning model requires higher computational resources and a large labeled dataset to get better accuracy. Unfortunately, it takes a lot of time and cost to collect and annotate the dataset because each utterance might consist of multiple or ambiguous emotions. Therefore, either traditional or advanced data augmentation methods have been applied to generate and synthesize randomly the training data samples~\cite{devries2017dataset}--\cite{qian2019data}. These studies only focused on using either GAN from feature space or adding noise/oversampling technique to generate and balance data samples, however, they are limited with sequence model and data space approaches.

In this article, we propose the HDA methods that combine traditional and GAN-based methods to generate and balance the labeled dataset. Then, the 3D log MelSpec low-level features are extracted as the inputs for the deep dilated convolutional-recurrent neural networks (DCRNN). The deep DCRNN learn and extract the high-level representations that are then fed into an attention layer to exploit the utterance-level features. Finally, we combine the softmax loss and center loss to classify the emotions from speech. We use the EmoDB for both the HDA state and emotion recognition state. The main contributions of this article are listed below:
\begin{itemize}
	\item We utilize the HDA methods that consist of time shifting, pitch shifting, WaveGAN, and SpecGAN to generate and balance samples on the EmoDB and IEMOCAP dataset.
	\item After that, we implement the deep ADCRNN to learn and extract utterance-level features from the generated 3D log MelSpec low-level features. 
	\item Then, the loss function is reconfigured by combining the softmax and center loss to classify the emotional speech from both original and augmented data.
	\item Experimental results prove that our proposed methods are better accuracy than the state-of-the-art methods on the EmoDB with the HDA methods.
\end{itemize}

This section aims to introduce the SER, the previous approaches to extract features and classify the emotion for the SER, and its applications. The literature research is presented in Section~\ref{sect2}. The proposed methodologies are addressed in detail in Section~\ref{sect3}. In Section~\ref{sect4}, the experimental results and comparison are presented and analyzed. We conclude this study and propose some potential future works in Section~\ref{sect5}.  

\section{Related Work} \label{sect2}

\subsection{Feature Extraction and Selection for SER}
Over the last decade, feature extraction and selection have been important parts of the SER. Researchers have tried to figure out the optimal and robust features, but they are challenging with which features should be chosen. Koduru \emph{et al.} used different type of features to improve the SER rate, such as pitch, energy, MFCC, ZCR, and discrete wavelet transform (DWT)~\cite{koduru2020feature}. Lamiaa Abdel-Hamid used prosodic, spectral and wavelet features that consist of the pitch, intensity, formants, MFCC, long-term average spectrum (LTAS), and wavelet to investigate for the SER~\cite{ABDELHAMID202019}. Atalay \emph{et al.} compared the feature selection techniques with MFCC features that include autoencoder, Chi-Square, and relief-F for emotion recognition in voice~\cite{atalay2018comparison}. Chen \emph{et al.} proposed a two-layer fuzzy multiple random forests (TLFMRF) algorithm to classify the emotion from extracted features that fuse from personalize and non-personalized features and separate into emotional classes by fuzzy C-means clustering technique~\cite{chen2020two}. Huang \emph{et al.} proposed a feature extraction method upon wavelet packet (WP) filterbank for the SER that outperforms the MFCC features and can be used for 2D facial emotion recognition (FER) and audio-visual bimodal emotion recognition system~\cite{huang2015extraction}. 

\subsection{Deep learning for SER}
Over the last decade, with the development of neural networks and deep learning, deep CNN and LSTM have been employed to extract features from the spectrogram representations of raw audio and classify emotions for the SER systems. Zhang \emph{et al.} investigated the deep CNN to extract 3D log MelSpec features, then designed a discriminant temporal pyramid matching (DTPM) strategy to concatenate the learned segment-level features, and used SVM classifier to recognize the emotions~\cite{zhang2017speech}. Tzirakis \emph{et al.} proposed an end-to-end multimodal that consists of CNN to extract speech features and a deep residual network of 50 layers (ResNet50) to extract visual features, then fed into two LSTM layers to extract the important features for the SER~\cite{tzirakis2017end}. Zhao \emph{et al.} designed 1D and 2D CNN with the LSTM for the SER that not only overcomes the shortcoming of the CNN and the LSTM but also takes advantage of the strength of them~\cite{zhao2019speech}. Sajjad \emph{et al.} proposed a method upon radial basis function network (RBFN) to clustering the key sequence segment, then all selected sequences are converted into spectrograms to extract features by CNN and learn the temporal information for classifying the emotions by bidirectional LSTM~\cite{sajjad2020clustering}. Yao \emph{et al.} investigated a fusion of 3 classifiers upon multi-task learning that consists of MelSpec combined with CNN (MS-CNN), low-level descriptors combined with recurrent neural networks (LLD-RNN), and hight-level statistical functions combined with deep neural network (HSF-DNN) for the SER~\cite{YAO202011}. Meng \emph{et al.} proposed a novel architecture for the SER using dilated CNN with residual block and bidirectional LSTM based on attention mechanism (ADRNN)~\cite{meng2019speech}. The ADRNN extracts the features and learn representation from 3D log MelSpec and then classifies the emotions using the loss function that applies the center loss together with the softmax loss.

\subsection{Attention mechanism for SER}
Since not all features equally contributed to recognizing the emotion from speech, recent studies have employed an attention mechanism for SER. Meng, Chen, and Xie \emph{et al.} employed the attention-based LSTM to learn the relevant high-level features representing for emotion states~\cite{chen20183},~\cite{meng2019speech, xie2019attention}. Yoon \emph{et al.} proposed a multi-hop attention mechanism for SER trained to calculate automatically the correlation between the modalities~\cite{yoon2019speech}. Peng \emph{et al.} proposed a sliding RNN method upon attention for SER that extracts the segment-level features and focuses only on the important emotional part of the speech features~\cite{peng2020speech}. Tarantino \emph{et al.} proposed a self-attention combined with a new global windowing system outperforming the previous state-of-the-art methods~\cite{tarantino2019self}. Zhao \emph{et al.} combined the bidirectional LSTM based on attention mechanism with a fully convolutional network based on the attention mechanism to extract deep spectrum representations for SER~\cite{zhao2019exploring}. Ho \emph{et al.} used the self-attention for RNN to exploit the context for each time step, then used the multi-head attention to fuse all representatives for predicting the emotions~\cite{ho2020multimodal}.

\subsection{Adversarial Data Augmentation for SER}
To deal with imbalanced data and reduce overfitting, researchers have been used the data augmentation method to generate or synthesize data samples. Huang \emph{et al.} proposed the data augmentation method for the training data by replacing the source data samples with the shorter overlapping samples extracted from them~\cite{huang2018multimodal}. Park \emph{et al.} proposed a SpecAugment method for speech recognition~\cite{Park2019}. SpecAugment includes features warping, frequency masking, and time masking that are applied to the inputs of a neural network. Rebai \emph{et al.} proposed a new DNN architecture taking advantage of both data augmentation and ensemble approaches to improve the accuracy of emotion recognition~\cite{rebai2017improving}. In recent years, GAN-based techniques have been developed to improve the accuracy of emotion recognition as a data augmentation method. Sahu \emph{et al.} used the applications of GAN to synthesizing features vectors for the SER that enhances the performance of classification~\cite{Sahu2018}. Yi \emph{et al.} proposed an adversarial data augmentation network (ADAN) that includes an autoencoder feature selection, a GAN, and an auxiliary classifier to improve the SER~\cite{yi2020impr}. The ADAN using the Wasserstein divergence instead of cross-entropy loss for training the GAN to generate feature vectors in both the original feature space and the latent space. Bao and Vu \emph{et al.} investigated a method upon Cycle consistent adversarial networks (CycleGAN) that transfers the feature vectors from a large speech corpus without labeled into synthetic features of emotion styles to improve classification performance~\cite{bao2019cyclegan}. Eskimez and Chatziagapi \emph{et al.} proposed a GAN method upon CNN to generate the spectrograms for training the SER model~\cite{chatziagapi2019data, eskimez2020gan}.

This study is motivated by the WaveGAN and SpecGAN in~\cite{DonahueMP19}, the works in~\cite{lent1989efficient, haghparast2007real}, the deep learning architecture for 3D log MelSpec in~\cite{meng2019speech}, and the loss function for speech emotion recognition combined contrastive-center (CT-C) loss with softmax loss proposed by Pham~\emph{et al.} \cite{pham2020method}. We combine these motivated approaches to conduct our work in the following aspects:
\begin{itemize}
	\item First, we apply and implement the WaveGAN, SpecGAN, pitch shifting, and time shifting as HDA methods to generate and synthesize training dataset.
	\item Second, the ADRNN is modified by removing all batch normalization (BN), then we also use a fully connected layer (FCN) with 64 units to obtain the reconfigured loss function.
	\item Third, we apply different loss functions, such as the softmax loss, the reconfigured softmax loss + center loss, the softmax loss + the center loss in \cite{meng2019speech}, and the CT-C loss + softmax loss in \cite{pham2020method} to validate the proposed method. We run several experiments to compare these loss functions with each other and with the previous works.
\end{itemize}

\section{Proposed Methodologies} \label{sect3}
\begin{figure*}[h!]
	\centering
	\includegraphics[width=0.95\linewidth]{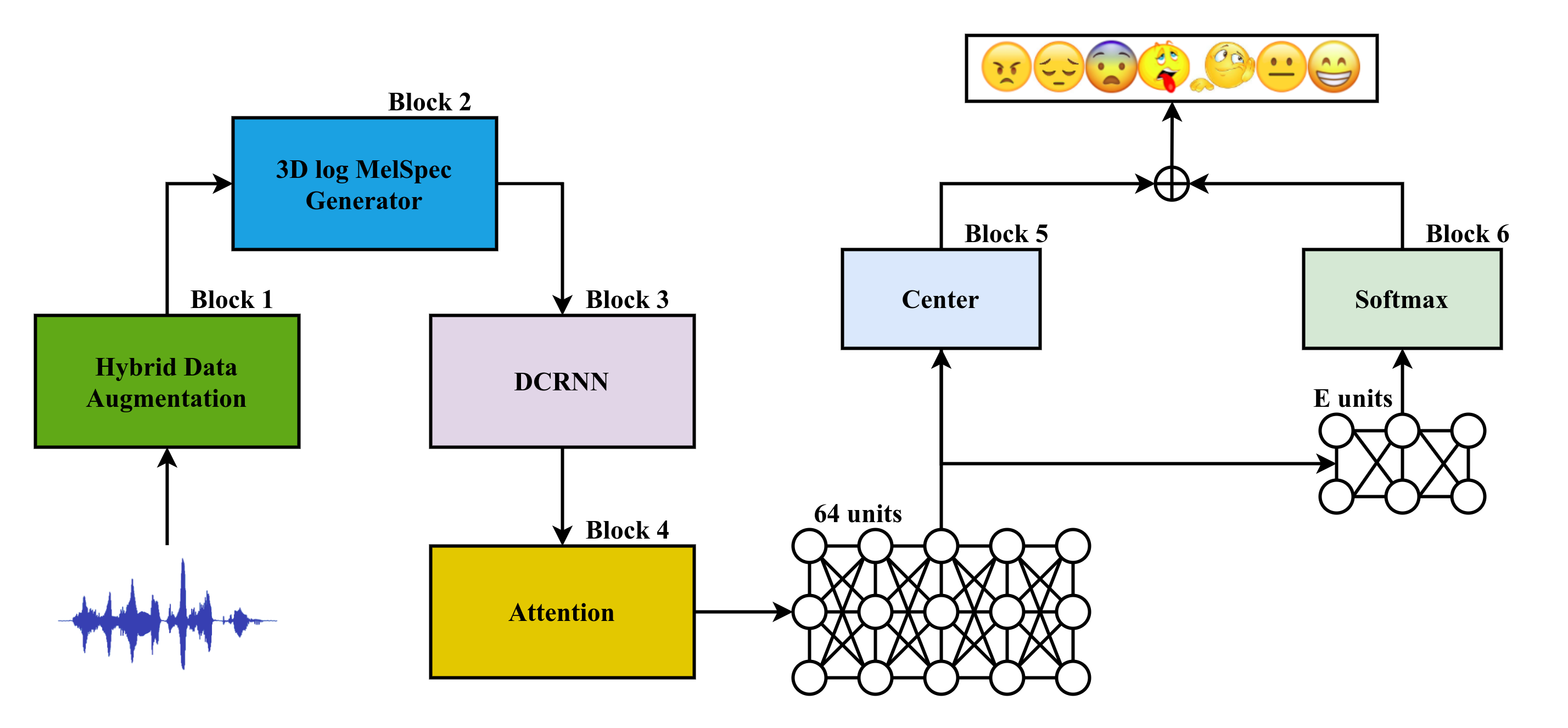}
	\caption{The whole network architecture.} \label{fig1}
\end{figure*}

Since the distribution of emotions in almost benchmark datasets and natural speech signals are not balance and lack of data. In this section, we utilize a baseline architecture for the SER system that deals with these problems and improves the recognition rate of emotions. The baseline architecture is shown in Fig.~\ref{fig1} consisting of six main blocks: Hybrid Data Augmentation block, 3D log MelSpec Generator block, DCRNN block, Attention block, Center block, and Softmax block. Table~\ref{tab1} shows the notations and their corresponding description used in this study.
\begin{table}[h!]
	\centering
	\caption{The notations and their description used in the proposed method.}\label{tab1}
	\begin{tabular}{c|l}
		\hline
		\textbf{Notations} &  \textbf{Description} \\
		\hline
		Block 1 & Hybrid data augmentation block \\
		Block 2 & 3D log Mel-spectrogram generator block \\
		Block 3 & Deep DCRNN architecture block \\
		Block 4 & Attention layer block \\
		Block 5 & Center loss block \\
		Block 6 & Softmax loss block \\
		\hline
	\end{tabular}
\end{table}

In the baseline architecture, we design the deep ADCRNN to learn and extract high-level representations from 3D log MelSpec low-level features for the SER. The deep ADCRNN is based on the ADRNN, but we modify it a bit by removing all BN layers after the dilated CNN layers. To overcome the challenging in RNN, such as complex dependencies, vanishing, and exploding gradients, we proposed dilated LSTMs to replace the BiLSTM. Furthermore, we use an FCN layer with 64 units to compute the center loss before the down-sampling shape to $E$ classes to compute the softmax loss. This work is quite different from the ADRNN because the center loss and the softmax loss in the ADRNN are computed after down-sampling shape to $E$ classes. The baseline architecture in Fig.~\ref{fig1} is designed as follows:
\begin{itemize}
	\item First, the speech signals are augmented to generate and balance data by the hybrid data augmentation.
	\item Second, we use a CNN layer to perform on 3D log MelSpec low-level features extracted by the 3D log MelSpec generator. 
	\item Third, we add 3 dilated CNN layers with residual block to extract temporal features. 
	\item Next, all feature maps are the input for the bidirectional LSTM to learn sequential features. 
	\item Then, we add an attention layer to exploit the utterance-level features from sequential features. 
	\item Finally, a loss function is used to classify the emotion by combining the softmax loss and center loss for the SER. 
\end{itemize}

The structure of this section is organized as follows: The hybrid data augmentation methods with two approaches are presented in Subsection~\ref{sub1}; the 3D log MelSpec extraction is described in Subsection~\ref{sub2}; and in Subsection~\ref{sub3}, we present a deep learning framework included the deep DCRNN architecture, attention layer, and the loss function to construct the baseline architecture for the SER system.

\subsection{Hybrid Data Augmentation (HDA) Methods} \label{sub1}
\subsubsection{Traditional Approaches}
\paragraph{Time Shifting} \text{}

Given a signal $\omega(t)$, we can shift the wave of the signal forward or backward by adding or subtracting a finite time $\tau$, respectively. The output $\chi(t)$ after shifting is defined as follows:
\begin{equation}
	\chi(t) = \omega(t \pm \tau),
\end{equation}
where $\tau = sr/100$ and the $sr$ is the sampling rate of the signal in this study.

\begin{figure}[h!]
	\centering
	\includegraphics[width=0.493\linewidth]{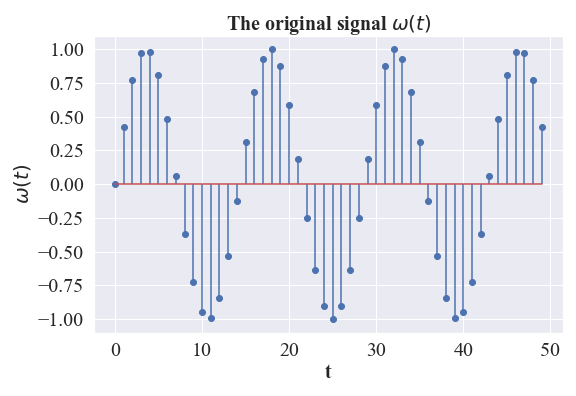}
	\includegraphics[width=0.493\linewidth]{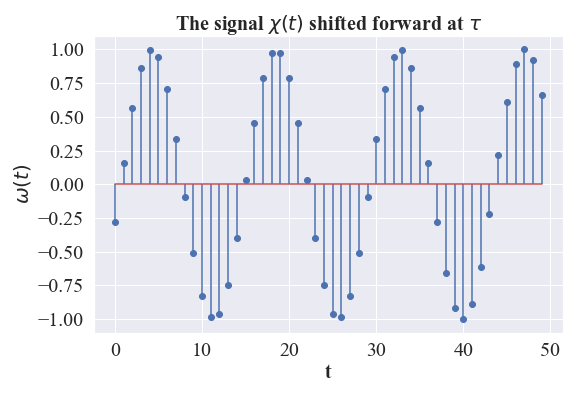}
	\includegraphics[width=0.493\linewidth]{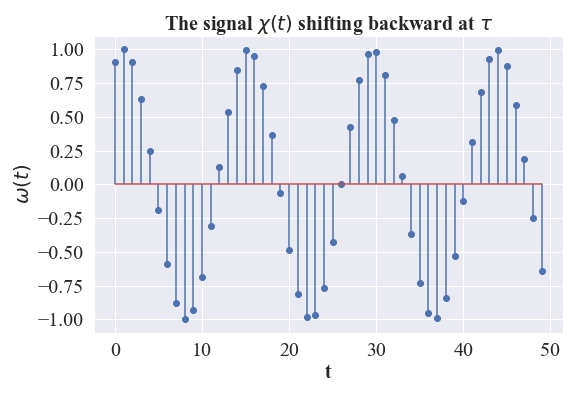}
	\includegraphics[width=0.493\linewidth]{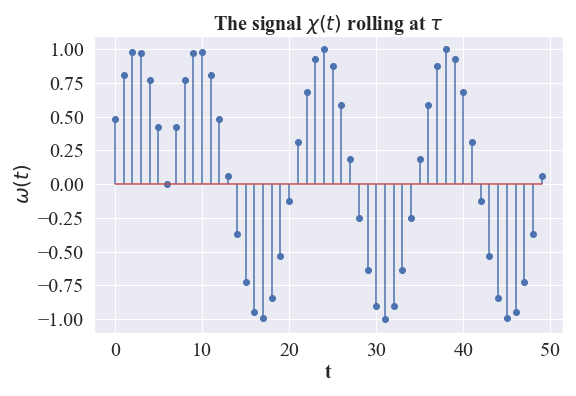}
	\caption{Examples of the time shifting and rolling.}
	\label{fig2}
\end{figure}

As using the time shifting, the signal is only shifted the position forward or backward without changing its amplitude. However, in this study, we not only want to shift the signal along its time but also want to roll it. Fig.~\ref{fig2} describes the examples of the time shifting and rolling in detail.

\paragraph{Pitch Shifting} \text{}

\begin{figure}
	\includegraphics[width=0.97\linewidth]{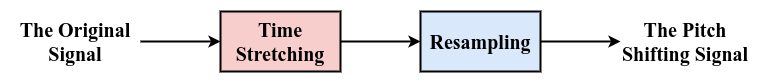}
	\caption{The process of the pitch shifting.}
	\label{fig3}
\end{figure}

Pitch shifting is the efficient algorithm proposed by Lent in \cite{lent1989efficient}. This algorithm was based on the time stretching and resampling methods~\cite{haghparast2007real}. In this study, pitch shifting is presented as in Fig.~\ref{fig3}.

\begin{figure}[h!]
	\centering
	\includegraphics[width=0.493\linewidth]{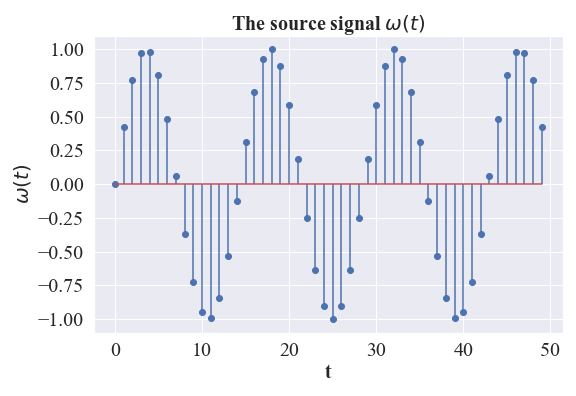}
	\includegraphics[width=0.493\linewidth]{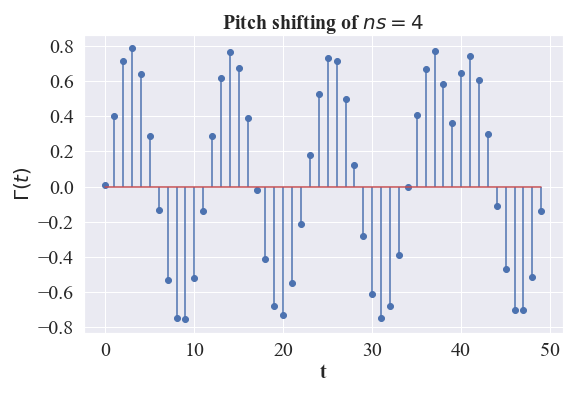}
	\includegraphics[width=0.493\linewidth]{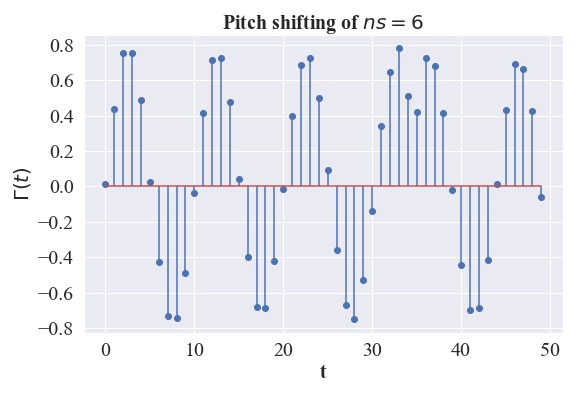}
	\includegraphics[width=0.493\linewidth]{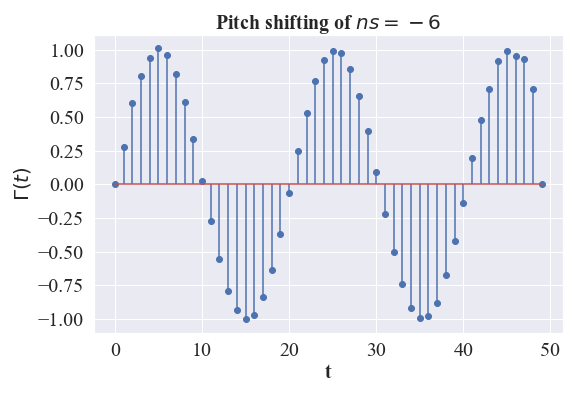}
	\caption{Examples of the pitch shifting.}
	\label{fig4}
\end{figure}

Given the number of half-steps $nhs$ and the number of bins $nbins$ in each octave, the time stretching is obtained by computing the time stretching ratio $S\_ratio$ as follows:
\begin{equation}
	S\_ratio = 2^{-\frac{float(nhs)}{nbins}}.
\end{equation}
The resampling is obtained by computing the resampling ratio $R\_ratio$ as follows:
\begin{equation}
	R\_ratio = \frac{T\_sr}{S\_sr},
\end{equation}
where $T\_sr$ and $S\_sr$ are the sampling rate of the target signal and the sampling rate of the source signal, respectively. If the $R\_ratio > 1$, then the pitch shifting signal is sped up, otherwise, it is slowed down. Fig.~\ref{fig4} describes the examples of the pitch shifting $\Gamma(t)$ of the source signal $\omega(t)$ using Librosa library in detail.

~\\

\subsubsection{GAN-based Approaches}
\paragraph{Brief Introduction of WaveGAN} \text{}

WaveGAN is the first GAN architecture for unsupervised audio synthesizing proposed by Donahue \emph{et al.}~\cite{DonahueMP19}. It is based on the novel deep convolutional GAN architecture for generating images (DCGAN) proposed by Radford \emph{et al.}~\cite{RadfordMC15}. The WaveGAN is constructed by modifying the DCGAN as follows:
\begin{itemize}
	\item Changing all the properties of the DCGAN model to support audio waves instead of image data;
	\item Using the 1D filters with the length of 25 instead of $5 \times 5$ 2D filters;
	\item Increasing and using the stride of 4 instead of $2 \times 2$ stride;
	\item Without using the BN in both generator and discriminator;
	\item Using the Wasserstein GAN and the gradient penalty (WGAN-GP) proposed by Gulrajani \emph{et al.}~\cite{NIPS2017_Gulrajani} to achieve Lipschitz continuity during the training process;
	\item Only using the phase shuffle operation in the discriminator that perturbs randomly the phase of each layer by $[-n,~n]$ samples, where $n$ is the hyperparameter.
\end{itemize}

The transposed convolution operation in the generator is up-sampling while it is down-sampling in the discriminator. Although the WaveGAN is a little bit modified, it still has the same number of parameters and the dimensional output is the same as the DCGAN. The parameters of the WaveGAN are described in detail in Table~\ref{tab2}. In which, $z, S, n\_ch$, and the $L\_out$ are the uniform distribution of 100 dimensions, the length of slice, the number of channels, and the linear output of the WaveGAN, respectively.


\begin{table*}[h!]
	\centering
	\caption{Description of the WaveGAN architecture~\cite{DonahueMP19}.}\label{tab2}
	\begin{tabular}{cccccc}
		\hline
		\multicolumn{6}{c}{\textbf{WaveGAN Architecture}}                                                                                                                                                                                                           \\ \hline
		\multicolumn{3}{c|}{\textbf{Generator}}                                                                                                   & \multicolumn{3}{c}{\textbf{Discriminator}}                                                                     \\ \hline
		\multicolumn{1}{c|}{\textbf{Operation}}          & \multicolumn{1}{c|}{\textbf{Kernel Size}} & \multicolumn{1}{c|}{\textbf{Output Shape}} & \multicolumn{1}{c|}{\textbf{Operation}}     & \multicolumn{1}{c|}{\textbf{Kernel Size}} & \textbf{Output Shape} \\ \hline
		\multicolumn{1}{c|}{Input z $\sim$Uniform(-1,1)} & \multicolumn{1}{c|}{-----}                     & \multicolumn{1}{c|}{(Bs, 100)}             & \multicolumn{1}{c|}{Input x or G(z)}        & \multicolumn{1}{c|}{-----}                     & (Bs, 16384, C)        \\ \hline
		\multicolumn{1}{c|}{Dense}                       & \multicolumn{1}{c|}{(100, 256D)}          & \multicolumn{1}{c|}{(Bs, 256D)}            & \multicolumn{1}{c|}{Conv1D 1 (S=4)}         & \multicolumn{1}{c|}{(25, C, D)}           & (Bs, 4096, D)         \\ \hline
		\multicolumn{1}{c|}{Reshape}                     & \multicolumn{1}{c|}{-----}                     & \multicolumn{1}{c|}{(Bs, 16, 16D)}         & \multicolumn{1}{c|}{LeakyReLU 1 (a=0.2)}    & \multicolumn{1}{c|}{-----}                     & (Bs, 4096, D)         \\ \hline
		\multicolumn{1}{c|}{ReLU 1}                      & \multicolumn{1}{c|}{-----}                     & \multicolumn{1}{c|}{(Bs, 16, 16D)}         & \multicolumn{1}{c|}{Phase Shuffle 1 (Bs=2)} & \multicolumn{1}{c|}{-----}                     & (Bs, 4096, D)         \\ \hline
		\multicolumn{1}{c|}{Transpose Conv1D 1 (S=4)}    & \multicolumn{1}{c|}{(25, 16D, 8D)}        & \multicolumn{1}{c|}{(Bs, 64, 8D)}          & \multicolumn{1}{c|}{Conv1D 2 (S=4)}         & \multicolumn{1}{c|}{(25, D, 2D)}          & (Bs, 1024, 2D)        \\ \hline
		\multicolumn{1}{c|}{ReLU 2}                      & \multicolumn{1}{c|}{-----}                     & \multicolumn{1}{c|}{(Bs, 64, 8D)}          & \multicolumn{1}{c|}{LeakyReLU 2 (a=0.2)}    & \multicolumn{1}{c|}{-----}                     & (Bs, 1024, 2D)        \\ \hline
		\multicolumn{1}{c|}{Transpose Conv1D 2 (S=4)}    & \multicolumn{1}{c|}{(25, 8D, 4D)}         & \multicolumn{1}{c|}{(Bs, 256, 4D)}         & \multicolumn{1}{c|}{Phase Shuffle 2 (Bs=2)} & \multicolumn{1}{c|}{-----}                     & (Bs, 1024, 2D)        \\ \hline
		\multicolumn{1}{c|}{ReLU 3}                      & \multicolumn{1}{c|}{-----}                     & \multicolumn{1}{c|}{(Bs, 256, 4D)}         & \multicolumn{1}{c|}{Conv1D 3 (S=4)}         & \multicolumn{1}{c|}{(25, 2D, 4D)}         & (Bs, 256, 4D)         \\ \hline
		\multicolumn{1}{c|}{Transpose Conv1D 3 (S=4)}  & \multicolumn{1}{c|}{(25, 4D, 2D)}         & \multicolumn{1}{c|}{(Bs, 1024, 2D)}        & \multicolumn{1}{c|}{LeakyReLU 3 (a=0.2)}    & \multicolumn{1}{c|}{-----}                     & (Bs, 256, 4D)         \\ \hline
		\multicolumn{1}{c|}{ReLU 4}                      & \multicolumn{1}{c|}{-----}                     & \multicolumn{1}{c|}{(Bs, 1024, 2D)}        & \multicolumn{1}{c|}{Phase Shuffle 3 (Bs=2)} & \multicolumn{1}{c|}{-----}                     & (Bs, 256, 4D)         \\ \hline
		\multicolumn{1}{c|}{Transpose Conv1D 4 (S=4)}    & \multicolumn{1}{c|}{(25, 2D, D)}          & \multicolumn{1}{c|}{(Bs, 4096, D)}         & \multicolumn{1}{c|}{Conv1D 4 (S=4)}         & \multicolumn{1}{c|}{(25, 4D, 8D)}         & (Bs, 64, 8D)          \\ \hline
		\multicolumn{1}{c|}{ReLU 5}                      & \multicolumn{1}{c|}{-----}                     & \multicolumn{1}{c|}{(Bs, 4096, D)}         & \multicolumn{1}{c|}{LeakyReLU 4 (a=0.2)}    & \multicolumn{1}{c|}{-----}                     & (Bs, 64, 8D)          \\ \hline
		\multicolumn{1}{c|}{Transpose Conv1D 5 (S=4)}    & \multicolumn{1}{c|}{(25, D, C)}           & \multicolumn{1}{c|}{(Bs, 16384, C)}        & \multicolumn{1}{c|}{Phase Shuffle 4 (Bs=2)} & \multicolumn{1}{c|}{-----}                     & (Bs, 64, 8D)          \\ \hline
		\multicolumn{1}{c|}{Tanh}                        & \multicolumn{1}{c|}{-----}                     & \multicolumn{1}{c|}{(Bs, 16384, C)}        & \multicolumn{1}{c|}{Conv1D 5 (S=4)}         & \multicolumn{1}{c|}{(25, 8D, 16D)}        & (Bs, 16, 16D)         \\ \hline
		&                                           & \multicolumn{1}{c|}{}                      & \multicolumn{1}{c|}{LeakyReLU 5 (a=0.2)}    & \multicolumn{1}{c|}{-----}                     & (Bs, 16, 16D)         \\ \cline{4-6} 
		&                                           & \multicolumn{1}{c|}{}                      & \multicolumn{1}{c|}{Reshape}                & \multicolumn{1}{c|}{-----}                     & (Bs, 256D)            \\ \cline{4-6} 
		&                                           & \multicolumn{1}{c|}{}                      & \multicolumn{1}{c|}{Dense}                  & \multicolumn{1}{c|}{(256D, 1)}            & (Bs, 1)               \\ \cline{4-6} 
	\end{tabular}
\end{table*}

~\\

\paragraph{Brief Introduction of SpecGAN} \text{}

SpecGAN is also proposed in \cite{DonahueMP19} that generates semi-invertible spectrograms to reconstruct frequency-domain audio or waveform. The SpecGAN is designed as follows:
\begin{itemize}
	\item First, the frequency-domain audio is converted to the spectrograms by obtaining the short-time Fourier transform (STFT) with the length of windows of 16 ms, the overlap between the successive windows of 8 ms, and the fast Fourier transform (FFT) size of 128. The spectrograms are scaled logarithmically to get better alignment.
	\item Next, the spectrograms are clipped to three standard deviations and normalized to $[-1,~1]$ scale.
	\item Then, the DCGAN is applied to train and generate the spectrograms.
	\item Finally, the Griffin-Lim algorithm~\cite{griffin1984signal} is employed with 16 iterations to convert the generated spectrograms to audio samples and estimate the phase. The parameters of the SpecGAN are described in detail in Table~\ref{tab3}.
\end{itemize}


\begin{table*}[h!]
	\centering
	\caption{Description of the SpecGAN architecture~\cite{DonahueMP19}.}\label{tab3}
	\begin{tabular}{cccccc}
		\hline
		\multicolumn{6}{c}{\textbf{SpecGAN Architecture}}                                                                                                                                                                                                        \\ \hline
		\multicolumn{3}{c|}{\textbf{Generator}}                                                                                                   & \multicolumn{3}{c}{\textbf{Discriminator}}                                                                  \\ \hline
		\multicolumn{1}{c|}{\textbf{Operation}}          & \multicolumn{1}{c|}{\textbf{Kernel Size}} & \multicolumn{1}{c|}{\textbf{Output Shape}} & \multicolumn{1}{c|}{\textbf{Operation}}  & \multicolumn{1}{c|}{\textbf{Kernel Size}} & \textbf{Output Shape} \\ \hline
		\multicolumn{1}{c|}{Input z $\sim$Uniform(-1,1)} & \multicolumn{1}{c|}{-----}                     & \multicolumn{1}{c|}{(Bs, 100)}             & \multicolumn{1}{c|}{Input x or G(z)}     & \multicolumn{1}{c|}{-----}                     & (Bs, 128, 128, C)     \\ \hline
		\multicolumn{1}{c|}{Dense}                       & \multicolumn{1}{c|}{(100, 256D)}          & \multicolumn{1}{c|}{(Bs, 256D)}            & \multicolumn{1}{c|}{Conv2D 1 (S=2)}      & \multicolumn{1}{c|}{(5, 5, C, D)}         & (Bs, 64, 64, D)       \\ \hline
		\multicolumn{1}{c|}{Reshape}                     & \multicolumn{1}{c|}{-----}                     & \multicolumn{1}{c|}{(Bs, 4, 4, 16D)}       & \multicolumn{1}{c|}{LeakyReLU 1 (a=0.2)} & \multicolumn{1}{c|}{-----}                     & (Bs, 64, 64, D)       \\ \hline
		\multicolumn{1}{c|}{ReLU 1}                      & \multicolumn{1}{c|}{-----}                     & \multicolumn{1}{c|}{(Bs, 4, 4, 16D)}       & \multicolumn{1}{c|}{Conv2D 2 (S=2)}      & \multicolumn{1}{c|}{(5, 5, D, 2D)}        & (Bs, 32, 32, 2D)      \\ \hline
		\multicolumn{1}{c|}{Transpose Conv2D 1 (S=2)}    & \multicolumn{1}{c|}{(5, 5, 16D, 8D)}      & \multicolumn{1}{c|}{(Bs, 8, 8, 8D)}        & \multicolumn{1}{c|}{LeakyReLU 2 (a=0.2)} & \multicolumn{1}{c|}{-----}                     & (Bs, 32, 32, 2D)      \\ \hline
		\multicolumn{1}{c|}{ReLU 2}                      & \multicolumn{1}{c|}{-----}                     & \multicolumn{1}{c|}{(Bs, 8, 8, 8D)}        & \multicolumn{1}{c|}{Conv2D 3 (S=2)}      & \multicolumn{1}{c|}{(5, 5, 2D, 4D)}       & (Bs, 16, 16, 4D)      \\ \hline
		\multicolumn{1}{c|}{Transpose Conv2D 2 (S=2)}    & \multicolumn{1}{c|}{(5, 5, 8D, 4D)}       & \multicolumn{1}{c|}{(Bs, 16, 16, 4D)}      & \multicolumn{1}{c|}{LeakyReLU 3 (a=0.2)} & \multicolumn{1}{c|}{-----}                     & (Bs, 16, 16, 4D)      \\ \hline
		\multicolumn{1}{c|}{ReLU 3}                      & \multicolumn{1}{c|}{-----}                     & \multicolumn{1}{c|}{(Bs, 16, 16, 4D)}      & \multicolumn{1}{c|}{Conv2D 4 (S=2)}      & \multicolumn{1}{c|}{(5, 5, 4D, 8D)}       & (Bs, 8, 8, 8D)        \\ \hline
		\multicolumn{1}{c|}{Transpose Conv2D 3 (S=2)}    & \multicolumn{1}{c|}{(5, 5, 4D, 2D)}       & \multicolumn{1}{c|}{(Bs, 32, 32, 2D)}      & \multicolumn{1}{c|}{LeakyReLU 4 (a=0.2)} & \multicolumn{1}{c|}{-----}                     & (Bs, 8, 8, 8D)        \\ \hline
		\multicolumn{1}{c|}{ReLU 4}                      & \multicolumn{1}{c|}{-----}                     & \multicolumn{1}{c|}{(Bs, 32, 32, 2D)}      & \multicolumn{1}{c|}{Conv2D 5 (S=2)}      & \multicolumn{1}{c|}{(5, 5, 8D, 16D)}      & (Bs, 4, 4, 16D)       \\ \hline
		\multicolumn{1}{c|}{Transpose Conv2D 4 (S=2)}    & \multicolumn{1}{c|}{(5, 5, 2D, D)}        & \multicolumn{1}{c|}{(Bs, 64, 64, D)}       & \multicolumn{1}{c|}{LeakyReLU 5 (a=0.2)} & \multicolumn{1}{c|}{-----}                     & (Bs, 4, 4, 16D)       \\ \hline
		\multicolumn{1}{c|}{ReLU 5}                      & \multicolumn{1}{c|}{-----}                     & \multicolumn{1}{c|}{(Bs, 64, 64, D)}       & \multicolumn{1}{c|}{Reshape}             & \multicolumn{1}{c|}{-----}                     & (Bs, 256D)            \\ \hline
		\multicolumn{1}{c|}{Transpose Conv2D 5 (S=2)}    & \multicolumn{1}{c|}{(5, 5, D, C)}         & \multicolumn{1}{c|}{(Bs, 128, 128, C)}     & \multicolumn{1}{c|}{Dense}               & \multicolumn{1}{c|}{(256D, 1)}            & (Bs, 1)               \\ \hline
		\multicolumn{1}{c|}{Tanh}                        & \multicolumn{1}{c|}{-----}                     & \multicolumn{1}{c|}{(Bs, 128, 128, C)}     &                                          &                                           &                       \\ \cline{1-3}
	\end{tabular}
\end{table*}

\subsection{3D log MelSpec Extraction} \label{sub2}
In this study, we choose the 3D log MelSpec that are low-level features as the inputs for the deep ADCRNN model. The 3D log MelSpec low-level features consist of static, deltas, and delta-deltas coefficients are obtained as follows:
\begin{itemize}
	\item First, the audio samples are converted to the MelSpec by performing the STFT with the length of windows of 25 ms, the overlap between the successive windows of 10 ms, the number of filterbanks of 40, the frame rate of 16 kHz, resulting in 512 frequency bins (corresponding the FFT size of 512) with linear spaces from 300 Hz to 8 kHz.
	\item Next, the static coefficient is obtained by scaling logarithmically the MelSpec. 
	\item Then, the deltas coefficient is obtained by computing the derivative of the static coefficient.
	\item Finally, the delta-deltas coefficient is obtained by computing the derivative of the deltas coefficient.
\end{itemize}

\subsection{Deep Learning Framework} \label{sub3}
\subsubsection{Deep ADCRNN Architecture} \text{} 

\begin{figure}[h!]
	\centering
	\includegraphics[width=0.97\linewidth]{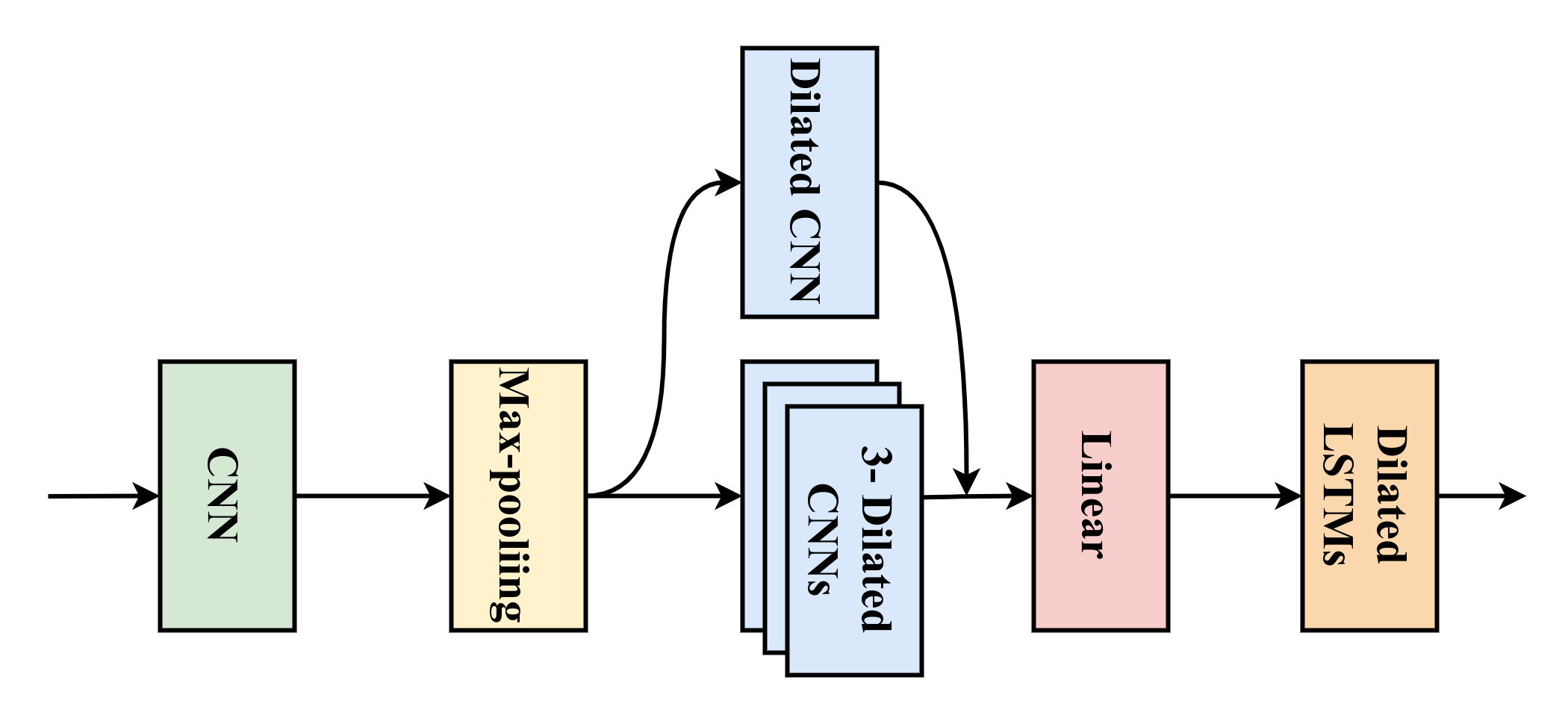}
	\caption{The deep DCRNN architecture.} \label{fig5}
\end{figure}

With the extracted 3D log MelSpec low-level features, the deep ADCRNN is used to learn and extract the high-level representation. The deep ADCRNN consists of 1 normal CNN layer, one max-pooling layer, 3 dilated CNN layers with skip dilated CNN connection, 1 linear layer, and 2 dilated LSTM layers. The first CNN layer has $3 \times 3$ kernel size, 128 feature maps, stride of 1, and \textit{valid} padding. Each dilated CNN layer has 256 feature maps with $3 \times 3$ kernel size, and \textit{same} padding. The dilation rate is set to 2 for the dilated CNN while it is set as list of (1, 2) for the dilated LSTM in this study. We only add the max-pooling layer after the first CNN layer to down-sample feature maps. The max-pooling layer has $2 \times 4$ kernel size, 128 feature maps, stride of $2 \times 4$, and \textit{valid} padding. To reduce the parameters effectively, we add a linear layer with 512 output units before fitting all feature maps into the dilated LSTMs. Each LSTM cell has 512 units and then we can obtain 512-dimensional sequential high-level representations. We also adopt a BN layer after the linear layer to improve the performance of the training process. The deep DCRNN architecture is shown in Fig.~\ref{fig5}.

\subsubsection{Attention-based Layer} \text{}

After extracting the high-level representations, an attention-based layer is added to exploit the utterance-level features for the SER because all sequential high-level representations do not contribute equally to represent the emotions from speech. In this study, the attention layer for the bidirectional LSTM is defined as follows:
\begin{equation}
	a_{\tau} = \sum_{\tau=1}^{T}\alpha_{\tau} \times h_{\tau},
\end{equation}  
where $a_{\tau}$ is the attention output, $h_{\tau} = \big[\overrightarrow{h_{\tau}};~\overleftarrow{h_{\tau}}\big]$ denotes the hidden state of the bidirectional LSTM output at time step $\tau$, $T$ is the total time-stamps, and the $\alpha_{\tau}$ is the normalized attention weight computed as below:
\begin{equation}
	\alpha_{\tau} = \frac{\exp\big(Z \cdot h_{\tau}\big)}{\sum\limits_{j=1}^{T} \exp \big(Z \cdot h_{\tau}\big)},
\end{equation}
where ($\cdot$) denotes the element-wise product and the $Z$ is the trainable weights.

Finally, we add the FCN layer with 64 output units that are used to compute the center loss and help the softmax loss to easier map the utterance-level features into $E$ different emotional classes as $E$ spaces. Only one dropout is applied after the FCN layer. Especially, this work is quite different from the previous work in \cite{meng2019speech} because the center loss is computed with 64 units instead of from the FCN with $E$ units corresponding $E$ classes. The center loss and the softmax loss are defined to compute the loss function are presented in detail in the Subsubsection~\ref{loss}.

\subsubsection{Loss Function} \label{loss} \text{} 

In this study, for the classification task, we combine the softmax loss and center loss as loss function to classify the emotion from speech and update weights during the training cycle. Because we want to both separate the features and discriminate them to recognize the emotions from speech. Therefore, we try to maximize the distance between the classes by the softmax loss and minimize the distance within-class by the center loss to optimize the training process.

The softmax loss or softmax cross-entropy loss is used to classify the features and it is widely applied in multiple classification problems. It is defined as below:
\begin{equation}
	\mathcal{L}_{SM} = -\sum\limits_{n=1}^{bs}\log\Big(\frac{e^{Z_{y_{n}}^{T}\times x_{n}+b_{y_{n}}}}{\sum\limits_{m=1}^{E} e^{Z_{m}^{T}\times x_{n}+b_{m}}}\Big),
\end{equation}
where the $\mathcal{L}_{SM}$ is the softmax loss and the $bs$ is the batch size or the number of samples in mini-batch.

The center loss that computes the distance between the features and their corresponding class centroids is defined as follows:
\begin{equation}
	\mathcal{L}_{CT} = \frac{1}{2}\sum\limits_{n=1}^{bs}\big\vert\big\vert x_{n} - C_{y_{n}}\big\vert\big\vert_{2}^{2},
\end{equation}
where the $\mathcal{L}_{CT}$ is the center loss and the $C_{y_{n}}$ is the centroid of class that the n-th sample belongs to.

The loss function is defined by combing the softmax loss and center loss as in Eq.~8.
\begin{equation}
	\mathcal{L}_{T} = \epsilon\times\mathcal{L}_{CT}+\mathcal{L}_{SM},
\end{equation}
where the $\mathcal{L}_{T}$ is the total loss function and the $\epsilon \in (0,1)$ is the factor to balance between the center and softmax losses. If the $\epsilon=0$, the loss function becomes the softmax loss.

\section{Experimental Results and Comparison} \label{sect4}
\subsection{Dataset}
\subsubsection{EmoDB dataset}
\begin{figure}[h!]
	\centering
	\includegraphics[width=\linewidth]{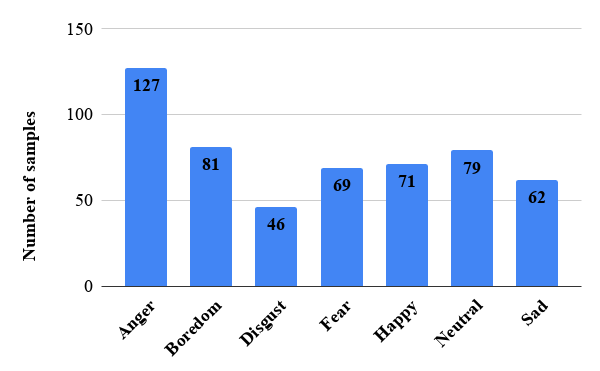}
	\caption{The detailed distribution of the EmoDB dataset.} \label{fig6}
\end{figure}

In this study, the Berlin Database of Emotional Speech (EmoDB) \cite{Burkhardt2005ADO} is used to implement the HDA methods and recognize the emotion from speech. The EmoDB consists of 535 audio data samples recorded by speaking the sentences in different emotions like happiness, sadness, anger, neutral, fear, disgust, and boredom. The speakers include 5 males and 5 females in the ages in a range of [25, 32]. The original database is recorded in 44.1 kHz and then resampled to 16 kHz. The EmoDB is visualized in detail in Fig.~\ref{fig6}.

%

\subsection{Experimental Setup}
The proposed model is trained on a single NVIDIA Geforce GTX 1050Ti 4 GB VRAM and 16 GB RAM with the TensorFlow framework. Our deep ADCRNN model is trained using Adam optimizer with learning rate $1e^{-4}$, batch size of 16, and the probability of every unit keep in dropout layer of 0.5. To get the robustness of training results, we also use 5-fold cross-validation to train the model.

The loss functions that are applied to train the classifiers in this study are surveyed as follows:
\begin{itemize}
	\item \textbf{$L_{f1}$:} Our reconfigured softmax loss and center loss;
	\item \textbf{$L_{f2}$:} Only the softmax loss;
	\item \textbf{$L_{f3}$:} The softmax loss and center loss in \cite{meng2019speech};
	\item \textbf{$L_{f4}$:} The softmax loss and CT-C loss in \cite{pham2020method}.
\end{itemize}

The HDA methods are applied to generate data for the bored, disgust, fear, happy, neutral, and sad emotions. We do experiments using $L_{f1}$ with 5 cases as follows:
\begin{itemize}
	\item Without using the HDA methods;
	\item Using the time shifting;
	\item Using the pitch shifting;
	\item Using the WaveGAN;
	\item Using the SpecGAN.
\end{itemize}

The Griffin-Lim algorithm has been developed in the Librosa library is used to reconstruct the audio samples. For 3D log MelSpec extraction, we use the framework \cite{lyons2020py} to extract the static, deltas, and delta-deltas coefficients of the log MelSpec.

\subsection{Results}
\subsubsection{Experiment without using the HDA Methods} \text{}

\begin{figure}[h!]
	\centering
	\includegraphics[width=\linewidth]{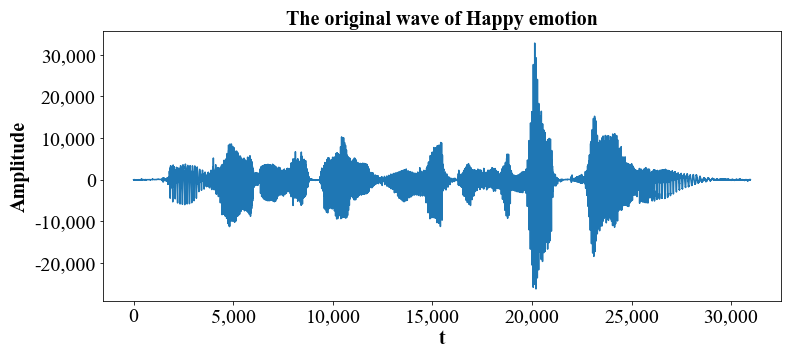}
	\includegraphics[width=\linewidth]{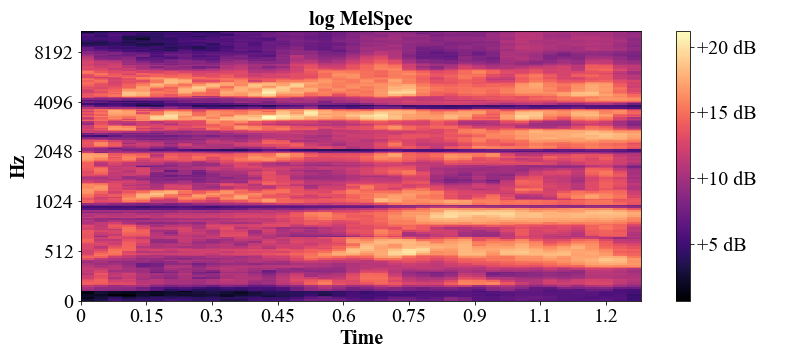}
	\caption{Visualization of the waveform and the corresponding log MelSpec of the original EmoDB dataset.}
	\label{fig7}
\end{figure}

Fig.~\ref{fig7} shows the waveform and the corresponding log MelSpec feature of the origin happy emotion on the EmoDB dataset.

\begin{figure}[htp]
	\centering
	\includegraphics[width=\linewidth]{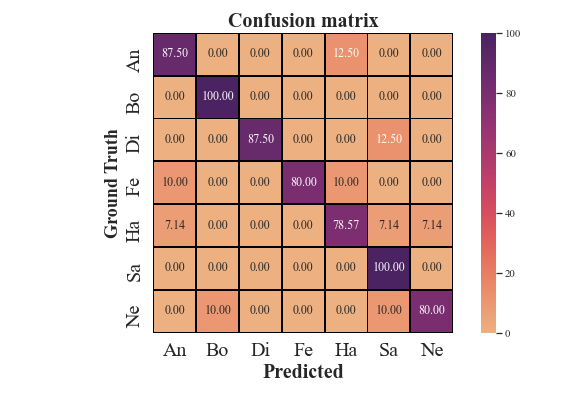}
	\caption{The experiment without using the HDA methods on the EmoDB.} \label{fig8}
\end{figure}

The confusion matrix of the experiment using the original dataset is shown in Fig.~\ref{fig8} that presents the predicted and ground truth emotions. The $An, Bo, Di, Fe, Ha, Sa$ and $Ne$ represent the angry, bored, disgust, fear, happy, sad, and neutral emotions, respectively. The accuracy of the bored, happy, disgust and sad emotions of our proposed model are better accurate than the ADRNN at 100.00\%, 78.57\%, 87.50\%, and 100.00\%, respectively. The proposed model can achieve notable improvement because the model extracts more features when computing the center loss after the FCN with 64 units.

\subsubsection{Experiment using the Time Shifting} \text{}

\begin{figure}[h!]
	\centering
	\includegraphics[width=\linewidth]{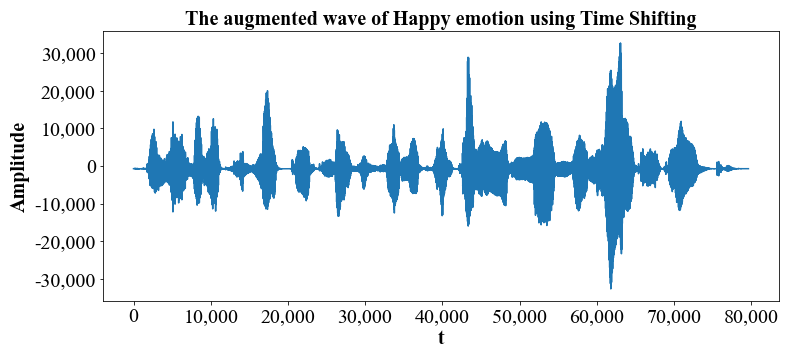}
	\includegraphics[width=\linewidth]{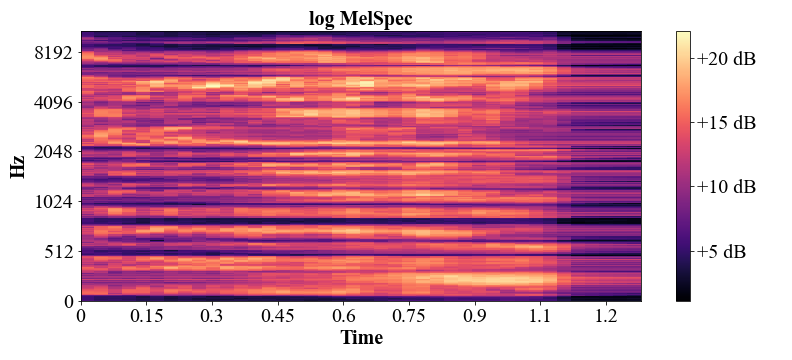}
	\caption{Visualization of the waveform and the corresponding log MelSpec of the time shifting EmoDB dataset.}
	\label{fig9}
\end{figure}

Fig.~\ref{fig9} shows the waveform and the corresponding log MelSpec feature of the time shifting happy emotion on the EmoDB dataset.

\begin{figure}[h!]
	\centering
	\includegraphics[width=\linewidth]{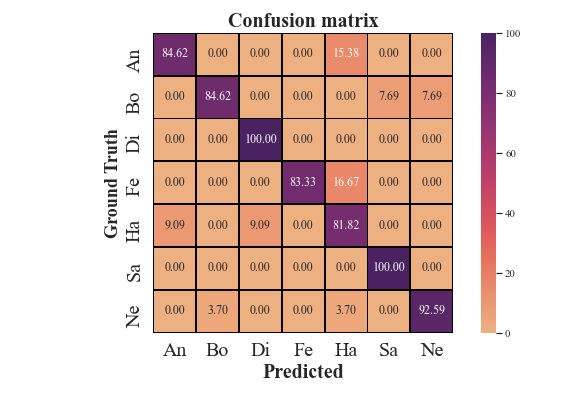}
	\caption{The experiment using the time shifting on the EmoDB.} \label{fig10}
\end{figure}

Fig.~\ref{fig10} describes the predicted and ground truth emotions of the experiment using time shifting. In the case of comparing with the ADRNN, our proposed model with time shifting augmentation achieve better accuracy in the disgust, happy, and sad emotions at 100.00\%, 81.82\%, and 100.00\%, respectively. The other emotions are a little less accurate than the ADRNN because of fewer data. On the other hand, the proposed model on the time shifting dataset is better than the original one in the disgust, fear, happiness, and neutral emotions at 100\%, 83.33\%, 81.82\%, and 92.59\%, respectively.

\subsubsection{Experiment using the Pitch Shifting} \text{}

\begin{figure}[h!]
	\centering
	\includegraphics[width=\linewidth]{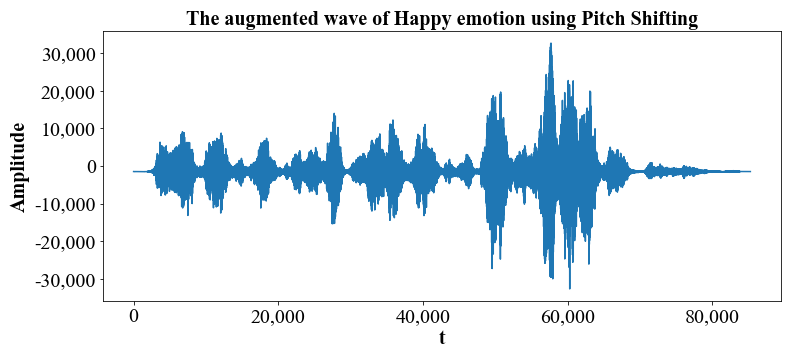}
	\includegraphics[width=\linewidth]{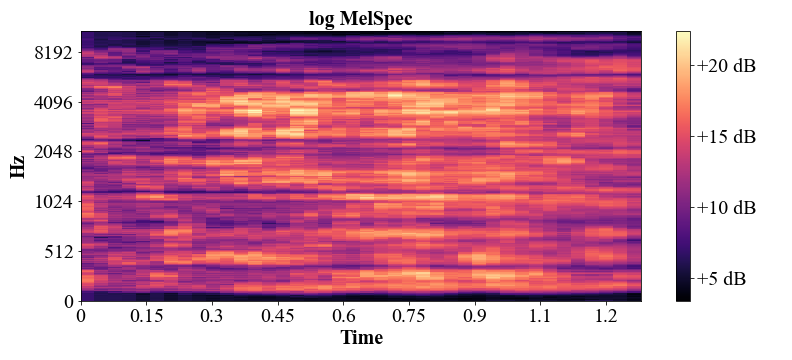}
	\caption{Visualization of the waveform and the corresponding log MelSpec of the pitch shifting EmoDB dataset.}
	\label{fig11}
\end{figure}

Fig.~\ref{fig11} shows the waveform and the corresponding log MelSpec feature of the pitch shifting happy emotion on the EmoDB dataset.

\begin{figure}[h!]
	\centering
	\includegraphics[width=\linewidth]{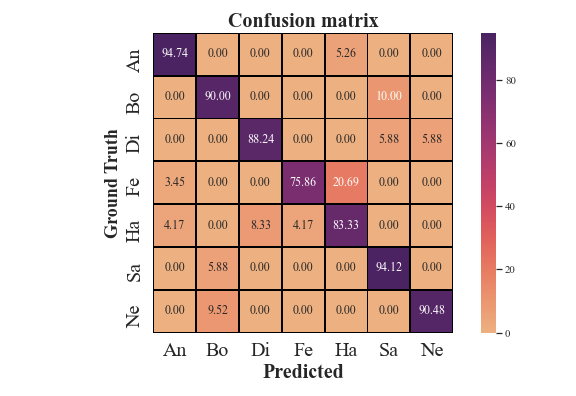}
	\caption{The experiment using the pitch shifting on the EmoDB.} \label{fig12}
\end{figure}

The confusion matrix of the predicted and ground truth emotion of the experiment using the pitch shifting on the EmoDB is shown in Fig.~\ref{fig12}. Using the pitch shifting, our proposed model achieves higher performance than the ADRNN of 1.83\%, 1.11\%, 5.63\%, 24.18\%, and 0.57\% notable improvement in the angry, bored, disgust, happy, and sad emotions. In the case of comparing with the original dataset, the model using the pitch shifting is better accurate in the angry, disgust, happy, and neutral emotions at 94.74\%, 88.24\%, 83.33\%, and 90.48\%, respectively.

\subsubsection{Experiment using the WaveGAN} \text{}

\begin{figure}[h!]
	\centering
	\includegraphics[width=\linewidth]{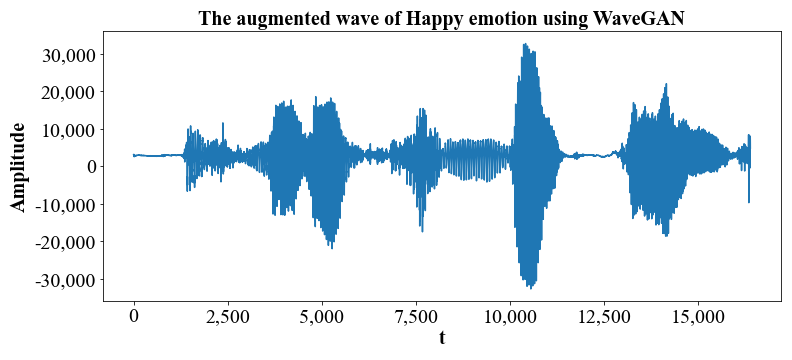}
	\includegraphics[width=\linewidth]{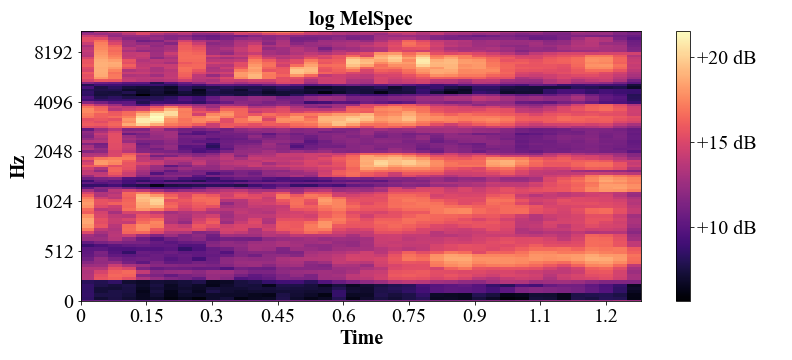}
	\caption{Visualization of the waveform and the corresponding log MelSpec of the WaveGAN EmoDB dataset.}
	\label{fig13}
\end{figure}

Fig.~\ref{fig13} shows the waveform and the corresponding log MelSpec feature of the WaveGAN happy emotion on the EmoDB dataset. In this experiment, the WaveGAN model has trained up to $90 \times 10^{3}$ steps to generate the augmented dataset.

\begin{figure}[h!]
	\centering
	\includegraphics[width=\linewidth]{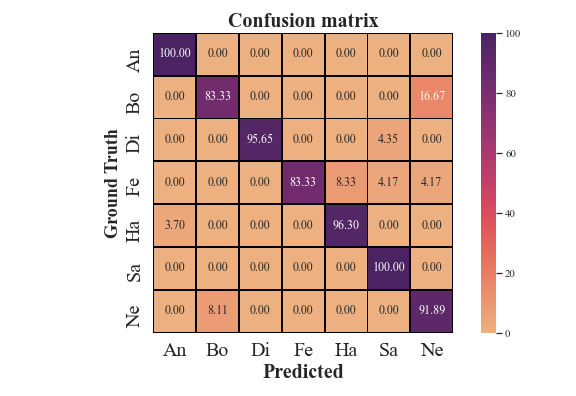}
	\caption{The experiment using the WaveGAN on the EmoDB.} \label{fig14}
\end{figure}

The confusion matrix of the emotion recognition rate for the experiment using WaveGAN on the EmoDB dataset is shown in Fig.~\ref{fig14}. In the angry, disgust, happy, and sad emotions, our proposed model could gain more improved accuracy than the ADRNN at 100.00\%, 95.65\%, 96.30\%, and 100.00\%, respectively. Especially, the accuracy in the happy emotion of the proposed model achieves higher than the ADRNN of 37.15\% notable improvement. Compare with the original dataset, the model using the WaveGAN augmentation is better accurate in the angry, disgust, fear, happy, and neutral emotions at 100\%, 95.65\%, 83.33\%, 96.30\%, and 91.89\%, respectively. The accuracy in the sad emotion is equal while another is less than the experiment using the original dataset.

\subsubsection{Experiment using the SpecGAN} \text{}

\begin{figure}[h!]
	\centering
	\includegraphics[width=\linewidth]{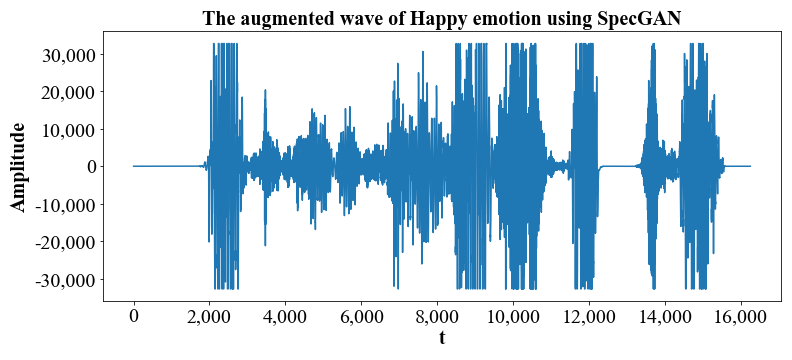}
	\includegraphics[width=\linewidth]{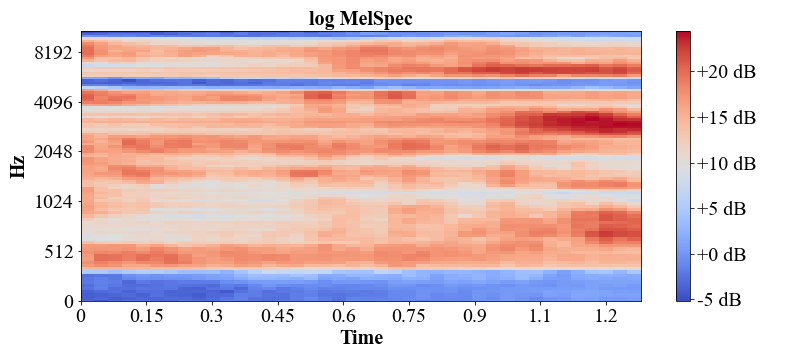}
	\caption{Visualization of the waveform and the corresponding log MelSpec of the SpecGAN EmoDB dataset.}
	\label{fig15}
\end{figure}

Fig.~\ref{fig15} shows the waveform and the corresponding log MelSpec feature of the SpecGAN happy emotion on the EmoDB dataset. The log MelSpec in this experiment is a little different from the others because SpecGAN is the spectral-based approach. The SpecGAN model has trained with $35 \times 10^{3}$ steps to generate the augmented SpecGAN dataset.

\begin{figure}[h!]
	\centering
	\includegraphics[width=\linewidth]{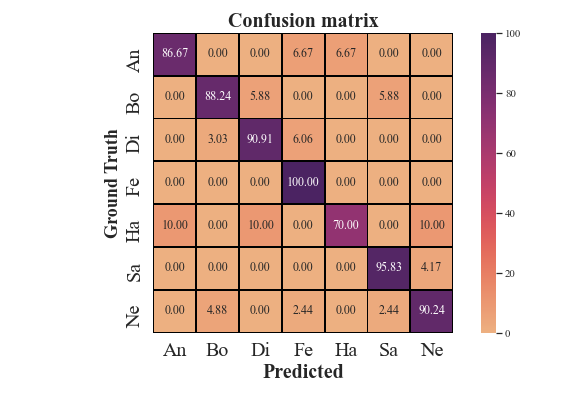}
	\caption{The experiment using the SpecGAN on the EmoDB.} \label{fig16}
\end{figure}

Fig.~\ref{fig16} shows the predicted and ground truth emotions of the experiment using SpecGAN on the EmoDB dataset in the confusion matrix. In the disgust, fear, happiness, and sad emotions, the proposed model using the SpecGAN achieves better accuracy at 90.91\%, 100.00\%, 70.00\%, and 95.83\% while the others are a little less than the ADRNN. In terms of comparing with the original data, the model using the SpecGAN gains more accuracy in the disgust, fear, and neutral emotions at 90.91\%, 100.00\%, and 90.24\% while the other emotions are less accurate. Especially, the accuracy in the fear emotion is better than the experiment using the original dataset of 20.00\% notable improvement.

\subsection{Comparison}
Table~\ref{tab4} shows the comparison of different loss functions with their corresponding mean and standard deviation of accuracy using the augmented dataset by the WaveGAN. The model that combines the deep DCRNN architecture (Block 3) with attention (Block 4) and $L_{f4}$ achieves the highest accuracy at 91.90\%. It proves that using the attention layer can gain higher accuracy than without using it while combining the softmax loss with the CT-C loss can achieve the highest accuracy. Besides, it also proves that the reconfigured loss function $L_{f1}$ outperforms the loss function $L_{f3}$ with 1\% notable improvement. Therefore, using 64 units to compute the center loss is more optimal than $E$ ones, where $E$ is the number of emotional states.

\begin{table}[h!]
	\centering
	\caption{The comparison of different models with their corresponding mean and standard deviation of accuracy using the WaveGAN data augmentation.}\label{tab4}
	\begin{tabular}{c|l|c}
		\hline
		\textbf{Model} & \textbf{Method} &  \textbf{Accuracy} (\%)\\
		\hline
		1 & Block 2 + Block 3 + $L_{f1}$ & 67.74 $\pm$ 5.89 \\
		2 & Block 2 + Block 3 + Block 4 + $L_{F2}$ & 84.87 $\pm$ 2.84 \\
		\textbf{3} & \textbf{Block 2 + Block 3 + Block 4 + $L_{f1}$} & \textbf{88.60 $\pm$ 2.98} \\
		4 & Block 2 + Block 3 + Block 4 + $L_{f3}$ & 87.57 $\pm$ 2.52 \\
		\textbf{5} & \textbf{Block 2 + Block 3 + Block 4 + $L_{f4}$} & \textbf{91.90 $\pm$ 0.86} \\
		\hline
	\end{tabular}
\end{table}

The models using the HDA methods are better accurate than the original dataset. Table~\ref{tab5} shows the comparison of the experimental results with their corresponding mean and standard deviation of accuracy using the reconfigured loss function $L_{f1}$. In which, the experiment using the WaveGAN achieves the highest accuracy at 88.47\%. With the GAN-based approach, the model using the WaveGAN is better than the SpecGAN while the model using pitch shifting is better than the time shifting in the traditional approach.

\begin{table}[h!]
	\centering
	\caption{The comparison of the experimental results with mean and standard deviation of accuracy using $L_{f1}$ as loss function.}\label{tab5}
	\begin{tabular}{c|l|c}
		\hline
		\textbf{Case} & \textbf{Method} &  \textbf{Accuracy} (\%)\\
		\hline
		1 & Without using the HDA methods &  85.66 $\pm$ 1.85 \\
		2 & Using the time shifting &  86.84 $\pm$ 2.15 \\
		\textbf{3} & \textbf{Using the pitch shifting} & \textbf{87.12 $\pm$ 0.84} \\
		\textbf{4} & \textbf{Using the WaveGAN} & \textbf{88.47 $\pm$ 2.76} \\
		5 & Using the SpecGAN & 87.32 $\pm$ 1.23 \\
		\hline
	\end{tabular}
\end{table}

The comparison of the proposed method with the previous works is shown in Table~\ref{tab6}. Our deep learning framework with the loss functions $L_{f1}$ and $L_{f4}$ outperform the ACRNN in~\cite{chen20183} and the ADRNN in~\cite{meng2019speech} at 88.60\% and 91.90\%, respectively. It proves that using the HDA methods with the proposed deep learning framework not only deals with the imbalanced and lack of data but also improves the recognition rate of emotional states. Simultaneously, it also proves that using the deep ADCRNN and the reconfigured loss function is better accuracy than the ADRNN with 3.21\% notable improvement.

\begin{table}[h!]
	\centering
	\caption{The comparison of the proposed method with the previous works.}\label{tab6}
	\begin{tabular}{l|c}
		\hline
		\textbf{Method} &  \textbf{Accuracy} (\%)\\
		\hline
		Chen and He \emph{et al.}~\cite{chen20183} & 82.82 \\
		Meng \emph{et al.}~\cite{meng2019speech} & 85.39 \\
		\textbf{Block 1 + Block 2 + Block 3 + Block 4 + $L_{f1}$} & \textbf{88.60} \\
		\textbf{Block 1 + Block 2 + Block 3 + Block 4 + $L_{f4}$} & \textbf{91.90} \\
		\hline
	\end{tabular}
\end{table}

\section{Conclusion} \label{sect5}
In this article, the HDA methods that combine both traditional and GAN-based approaches are proposed to generate and balance data for the SER. Besides, the deep ADCRNN is implemented to learn and extract the utterance-level features from 3D log MelSpec low-level ones. Furthermore, the loss function combining the softmax and center losses is investigated to improve the accuracy of emotion recognition. Experimental results prove that the HDA methods for speech emotion recognition can achieve better accuracy than the state-of-the-art methods in case of dealing with limited and imbalanced data.

Although the proposed hybrid augmentation methods and deep neural networks for the SER in this article achieves better accuracy and performance in terms of imbalance and lack of data, there are still a lot of aspects that can be improved and dived into research. In the future, we will investigate the multi-features fusion and multi-modal to exploit the robust and optimal features for the SER. Besides, we will also employ keyword spotting for the SER to integrate into the real-time systems.
\bibliographystyle{IEEEtran}
\end{document}